\documentclass[aps,prx,twocolumn,floats,showpacs,superscriptaddress,nofootinbib]{revtex4-1}
\usepackage{graphicx,epsfig}
\usepackage{times,bbm}
\usepackage{graphics,dcolumn,bm,float}
\usepackage{amssymb,amsmath,rotate,color}
\usepackage[title,titletoc,toc]{appendix}

\usepackage{booktabs}

\usepackage[pagebackref=false,colorlinks,linkcolor=magenta,citecolor=cyan,urlcolor=magenta]{hyperref}

	\newcommand{\be}{\begin{equation}}
	\newcommand{\ee}{\end{equation}}

	\newcommand{\ba}{\begin{align}}
	\newcommand{\ea}{\end{align}}

	\newcommand{\bearr}{\begin{eqnarray}}
	\newcommand{\eearr}{\end{eqnarray}}

	\newcommand{\bs}{\boldsymbol}
	\newcommand{\bmt}{\left[\begin{matrix}}
		\newcommand{\emt}{\end{matrix}\right]}

\begin{document}

\title{Fast nuclear spin relaxation rates in tilted cone Weyl semimetals:\\Redshift factors from Korringa relation}

\author{A. Mohajerani}
\email{a.mohajerani@modares.ac.ir}
\affiliation{Department of Basic Sciences$,$ Tarbiat Modares University (TMU)$,$ Tehran14115-175$,$ Iran}

\author{Z. Faraei}
\email{zahra.faraei@gmail.com}
\affiliation{Department of Physics$,$ Sharif University of  Technology$,$ Tehran 11155-9161$,$ Iran}

\author{S. A. Jafari}
\email{jafari@physics.sharif.edu}
\affiliation{Department of Physics$,$ Sharif University of  Technology$,$ Tehran 11155-9161$,$ Iran}

\date{\today}

\begin{abstract}
Spin lattice relaxation rate is investigated for 3D tilted cone Weyl semimetals (TCWSMs). The nuclear spin relaxation rate is presented as a function of 
temperature and tilt parameter. We find that the relaxation rate behaves as $(1-\zeta^2)^{-\alpha}$ with $\alpha\approx 9$ where $0\le \zeta < 1$ is the tilt parameter.
We demonstrate that such a strong enhancement for $\zeta\lesssim 1$ that gives rise to very fast relaxation rates, is contributed by the combined effect of a new 
hyperfine interactions arising from the tilt itself, and the anisotropy of the ellipsoidal Fermi surface. Extracting an effective density of states (DOS) $\tilde\rho$ 
from the Korringa relation, we show that it is related to the DOS $\rho$ of the tilted cone dispersion by the "redshift factor" $\tilde\rho=\rho/\sqrt{1-\zeta^2}$. 
We interpret this relation as NMR manifestation of an emergent underlying spacetime structure in TCWSMs. 
\end{abstract}


\maketitle
\narrowtext
\section{\label{sec:level1}Introduction}
Weyl semimetals (WSMs) such as TaAs~\cite{xu2015discovery,lv2015experimental,yang2015weyl}, NbAs~\cite{zheng2016atomic,liu2016evolution}, TaP~\cite{ xu2015experimental,xu2016observation}, NbP~\cite{belopolski2016criteria,souma2016direct,xu2017distinct} and YbMnBi$_2$~\cite{borisenko2019time} are materials realizations of the Weyl fermions in 
condensed matter physics in the sense that their low energy excitations are Weyl fermions and identified by Weyl equation~\cite{weyl1929elektron}. 
Whereas Weyl fermions have yet to be found as free particles in particle physics, condensed matter  is the only platform where they exist. 
In condensed matter realization of Weyl fermions, the right-handed and left-handed fermions (corresponding to those with their momentum parallel or anti-parallel to their spins) 
are organized around two different points in the Brillouin zone, and as such their Loretnz symmetry is broken~\cite{zheng2018quasiparticle}. 
Indeed, Weyl semimetals are non-degenerate analogs of Dirac semimetals which can be emerged by breaking either time-reversal or inversion symmetry in Dirac semimetals~\cite{wehling2014dirac,faraei2017superconducting}. WSMs feature pairs of Dirac cone type electronic band structure in their bulk and  Fermi arcs on the surface which lead 
to exotic  phenomena including negative magnetoresistance~\cite{son2013chiral}, nonlocal transport~\cite{parameswaran2014probing}, 
quantum anomalous Hall effect~\cite{yang2011quantum,yang2015chirality}, and unconventional superconductivity~\cite{cho2012superconductivity,wei2014odd,kim2016probing}.  

The Lorentz symmetry can be broken in more interesting ways in WSMs that can be interpreted as a new spacetime structure~\cite{Volovik2016,Nissinen2017,Tohid2019Spacetime,OjanenPRX,Ojanen2019,Sahar2019Polarization,Jafari2019,Bergholtz2020}. This deviation from Lorentz symmetry at the basic solid-state physics level, 
manifests as the deformation of cone shaped band structure resulting in tilted cone Weyl semimetals (TCWSMs) with point/hyperbolic Fermi surfaces 
(zero/nonzero density of states at the node) dubbed type-I and type-II WSMs respectively~\cite{Soluyanov2015}. 
The tilt deformation of Weyl equation can be formalized by an additional term in the Weyl Hamiltonian via an dimensionless tilt parameter 
$\boldsymbol{\zeta}=(\zeta_x, \zeta_y, \zeta_z)$ of magnitude $\zeta$ where $0\ \textless\ \zeta\ \textless\ 1$ and $\zeta\ \textgreater\ 1$ correspond to type-I and type-II respectively~\cite{Soluyanov2015,Tohid2019Spacetime,Sahar2019Polarization}. The tilt term of the Hamiltonian being proportional to the unit matrix in k-space, 
reshapes spherical Fermi surface to ellipsoidal one which results in increasing of the DOS at energies away from the Weyl node.
This is how the tilt deformation can alter the solid state properties of TCWSMs by modifying their energy levels. 

The nuclear magnetic resonance (NMR) as a powerful bulk spectroscopy exploits the week interaction between nuclear spins and surrounded electrons (Hyperfine interaction) to probe electronic properties of the materials~\cite{slichter2013principles}. Evidently, interaction between nuclear spins and spin angular momentum of quasiparticles have dominant contributions to the hyperfine interaction~\cite{alloul2015nmr}. Nevertheless, recent theoretical and experimental studies find that in Dirac and Weyl semimetals hyperfine interaction is different from the case of parabolic band structure which arises from the coupling of the spin and orbital degrees of freedom in linear band structure of Weyl/Dirac materials. In these systems the interaction between nuclear spins and electron orbital angular momentum comes into play and overwhelms the spin hyperfine interaction which leads to anomalous temperature dependence of spin lattice relaxation rate~\cite{okvatovity2016anomalous, yasuoka2017emergent, okvatovity2019nuclear, maebashi2019nuclear, yasuoka2017emergent}. 
Indeed, recent $^{13}$C NMR experiment on the quasi-two-dimensional organic conductor $\alpha-(BEDT-TTF)_{2}I_3$ has revealed that the 
local spin susceptibility and electron correlations are strongly angular dependent on the cone~\cite{hirata201113}.

What else can we learn from NMR reveal when the tilt is introduced to WSMs to make them TCWSMs?
The purpose of this work is to focus in the $k_BT\ll \mu$ limit of a TCWSm and to show that in the NMR spectroscopy of these materials, in addition to the
effects arising from deformation of the spherical Fermi surface to ellipsoidal Fermi surface (with eccentricity $\zeta$), there is a unique term arising
from the tilt of TCWSMs that generates its own coupling to nuclear spin degrees of freedom. Such a term has no analogue in the rest of solid state systems. 
In this study we theoretically illustrate how tilt parameter along with orbital magnetism, significantly contributes to the spin lattice relaxation rate in TCWSMs
and leads to a $(1-\zeta^2)^{-\alpha}$ dependence on the tilt parameter $\zeta$  with $\alpha\approx 9$, which strongly diverges for $\zeta\lesssim 1$. 
This suggests that the tilt can be considered as an additional relaxation channel in the electronic degree of freedom, through which nuclear spins relax back to the equilibrium. 
Indeed the DOS, $\rho$ is enhanced by a factor of $(1-\zeta^2)^{-2}$ whose square -- having Korringa relation in mind -- 
contributes an exponent $4$ to the total $\alpha\approx 9$.  
The tilt parameter further shows up explicitly in NMR rate via electron-spin hyperfine interaction. The orbital and tilt parts of the hyperfine
interaction couple to the nuclear spins in a same way and the tilt term as well as orbital term jointly contribute to the NMR rate. 
As such, it is not possible to separate the individual contributions of electron spin, orbital and tilt degrees of freedom. 
We show that the resulting complicated matrix elements will contribute another $4$ to the total exponent of $9$. Where does a missing exponent of $1$ come from?
Another purpose of this paper is to show that this discrepancy contains the physics of gravitational redshift. 

This paper is structured as follows: In section~\ref{formulation.sec} we formulate the hyperfine interaction in TCWSMs
and show how the presence of the tilt itself generates an additional terms in the hyperfine interaction giving rise to 
new relaxation mechanism. In section~\ref{Solution.sec} we apply our formulation to a TCWSM and find that the tilt enhances the 
relaxation rate via both a density of states (DOS) and an additional effect arising from the hyperfine interaction. We end the paper 
by a discussion of the redshift factors appearing in NMR measurements. 

\section{tilted Weyl fermions in the local field of nuclear spin}
\label{formulation.sec}
 The isotropic low energy effective Hamiltonian for a tilted Weyl semimetal is described by three dimensional Weyl equation 
of e.g. $+1$ chirality~\cite{goerbig2008tilted, trescher2015quantum},
 \begin{equation}
H(\textbf{k})=\hbar v_F[\boldsymbol{\sigma}\cdot \bs k+\boldsymbol{\zeta}\cdot\bs k ~\sigma_0],\label{eq:1}
  \end{equation}
where k is the momentum measured from the Weyl node, $\bs\sigma=(\sigma_1,\ \sigma_2,\ \sigma_3 )$ are Pauli matrices which present physical spin of quasiparticles and $\sigma_0$ denotes $2\times2$ identity matrix. $v_F$ and $\zeta$ are Fermi velocity and tilt parameter respectively. The eigenvalues of Eq.\eqref{eq:1} read
\be
\varepsilon_\pm(\bs k)=\hbar v_F(\bs\zeta\cdot\bs k \pm |\bs k|).\label{eq:2}
\ee
where $\pm$ denote the upper/lower bands touching at the Weyl nodes and correspond to the spinors eigenstates
$ |\bs k,\pm\rangle=\begin{pmatrix}
 k_x+ik_y &
 \pm k-k_z
\end{pmatrix}^\dagger $. The dimensionless tilt parameter $|\zeta|< 1$ and $\zeta\ > 1$ of magnitude $\zeta$ 
represent type-I and type II Weyl materials respectively. The tilt transforms the spherical Fermi surface of WSMs into ellipsoidal surfaces with eccentricity $\zeta$ 
and therefore  changes the the density of states per unit volume (DOS). The DOS in tilted Weyl materials at energy $\varepsilon$ above or below the Weyl node 
is given by $\frac{\varepsilon^2}{\pi^2 \hbar^3 v^3_F (1-\zeta^2)^2}$ which is trivially 
enhanced by a factor $(1-\zeta^2)^{-2}$ compared to non-tilted ones. 

To derive an expression for the relaxation rate in TCWSMs, we first investigate hyperfine interaction which has notable alteration due to  the tilt parameter. The  nuclear magnetic moment $\boldsymbol{\mu_n}=\hbar \gamma_n  \boldsymbol{I}$ with $\gamma_n$ nuclear gyromagnetic ratio and $\textbf{I}$ nuclear spin, induces a local magnetic field 
acting on each Weyl fermion. The tilted Weyl Hamiltonian  for an electron (charge -e) in the presence of nuclear field will be
\be 
  H(\bs k)=v_F \hbar \left[\bs\sigma \cdot  (\bs k + e \textbf{A}) + \bs \zeta \cdot (\bs k + e \textbf{A})\right]\ +\ g \mu_B\  \textbf S \cdot \textbf B,\label{eq:3}
\ee
where 
\be 
   \textbf{A}=-i\mu_0 \frac{\boldsymbol{\mu_n}\times\textbf{q}}{q^2},\label{eq:4}
\ee 
is the vector potential that leads to the local nucleus  magnetic field 
$\textbf{B}$, $g$ is the electron g-factor, $\mu_B$ is the Bohr magneton, and $\textbf{q}= \bs{k}^\prime-\bs{k}$. Hence, the  hyperfine interaction consists in 
three terms given by
\bearr
   &&H^{\rm spin}_{\rm hf}=\frac{g \hbar \mu_B \mu_0} {2}\  \boldsymbol{\mu_n}\cdot (\frac{\textbf{q}\times \textbf{q}\times \boldsymbol{\sigma}}{q^2})\label{eq:5},\\
   &&H^{\rm orbital}_{\rm hf}=-i e v_F \mu_0 \  \boldsymbol{\mu_n}\cdot \frac{\textbf{q}\times\boldsymbol{\sigma}}{q^2},\label{eq:6}\\
   &&H^{\rm tilt}_{\rm hf}=-i e v_F \mu_0 \  \boldsymbol{\mu_n}\cdot \frac{\textbf{q}\times\boldsymbol{\zeta}}{q^2}.\label{eq:7}
\eearr
Although the Eq.~\eqref{eq:5} and Eq.~\eqref{eq:6}, the interaction between nuclear spin, and spin and angular momentum of Weyl fermions, 
are well investigated in the NMR parameters of the Weyl and Dirac materials~\cite{okvatovity2019nuclear,maebashi2019nuclear}, 
but the  $H^{\rm tilt}_{\rm hf}$ is the new player in the nuclear relaxation rate of TCWSMs that is directly caused by the tilt parameter $\bs\zeta$. 
Since the tilt parameter in the Eq.~\eqref{eq:7} follows the same coupling pattern as orbital angular momentum in Eq.~\eqref{eq:6}, $H^{\rm tilt}_{\rm hf}$ 
is expected to have equally important contribution to the NMR rate. This is the essential conceptual point of this paper. 
Since we will be interested in highly doped TCWSMs where $k_BT \ll \mu$, the vector $\textbf{q}$ connecting two states on the Fermi surface
will not be small, and therefore the $H^{\rm spin}_{\rm hf}$ will have comparable contribution to the other two terms. So we keep all the temrs. 
Furthermore, ellipsoidal Fermi surface of tilted Weyl fermions will modify the natural coordinate system describing the Fermi surface (see the appendix)
whose Jacobian generates further implicit dependence on $\zeta$ that affects the matrix elements of
$H^{\rm spin}_{\rm hf}$ and $ H^{\rm orbital}_{\rm hf} $ terms.

 \section{Spin lattice relaxation rate for tilted weyl semimetals}
\label{Solution.sec}
The nuclear spin lattice relaxation time $T_1$ is determined by the part of the hyperfine interaction which nuclear spins flip to relax back to the equilibrium
denoted by $ H_{\rm hf}^\pm$. The NMR rate is given by the following fromula~\cite{slichter2013principles},
\begin{widetext}
\be
  \frac{1}{T_1 T} = \frac{\pi k_B}{\hbar} \int\frac{d\bs k'}{(2\pi)^3}\int\frac{d\bs k}{(2\pi)^3} \ |\langle \bs k',n'|H_{\rm hf}^-|\bs k ,n\rangle|^2\ 
  \left[-\frac{\partial f(\varepsilon)}{\partial\varepsilon}\right]\ \delta(\varepsilon-\varepsilon^\prime+\hbar\omega_0),
\ee
\end{widetext}
where $|\bs k\rangle$s and $|n\rangle$s denote quantum states of the electrons and nuclear spin, $f(\varepsilon)$ is Fermi Dirac distribution function, and $\omega_0$ is Larmor frequency. For a TCWSM the scattering matrix elements obtain by the following expression
\begin{widetext}
\be 
  \langle n^\prime \bs k'|H_{\rm hf}^-|\bs k n\rangle=\frac{\hbar e \mu_0 \gamma_n  F_N}{q^2} \left[ i v_F \langle \bs k'|(\textbf{q}\times \bs{\sigma})_-|\bs k\rangle+ i v_F \langle \bs k'|(\textbf{q}\times \bs{\zeta})_-|\bs k\rangle-\frac{\hbar}{2m_e}\langle \bs k'| (\textbf{q}\times \textbf{q}\times \boldsymbol{\sigma})_-|\bs k\rangle\right],\label{eq:9}
\ee
where $F_N=\langle n'|\textbf{I}^-|n\rangle$ is a constant coefficient for local isotropic interactions. In this paper, 
we consider the Fermi level in the upper branch of the Weyl node band far enough from Weyl node. For $\mu$ more than a fraction 
of an electron volt, even the room temperature satisfies $k_BT\ll \mu$ which allows for further simplifications of the 
Fermi-Dirac functions. Then only excitations in upper band are relevant. We further choose the momenta axis such that $\boldsymbol{\zeta}=(0, 0, \zeta)$ and the unit which in $\hbar=v_F=k_B=1$. The spin lattice relaxation rate in terms of the new coordinates can be rewritten as

\be
\frac{1}{T_1 T}=\frac{\mu_{0}^2\ \gamma_{n}^2\ e^2\  F_{N}^2}{65\  \pi^5}\int\frac{d\varepsilon\ d\Omega_k\ d\Omega_k^{\prime}}{(1+\zeta \cos\theta)^3 (1+\zeta \cos\theta^\prime)^3}\  h(\varepsilon,\Omega_k,\Omega_{k^\prime})\left[-\frac{df(\varepsilon)}{d\varepsilon}\right].\label{eq:10}
\ee
\end{widetext}
 
\begin{figure} [b]
   \centering 
   \includegraphics[width=.5\textwidth]{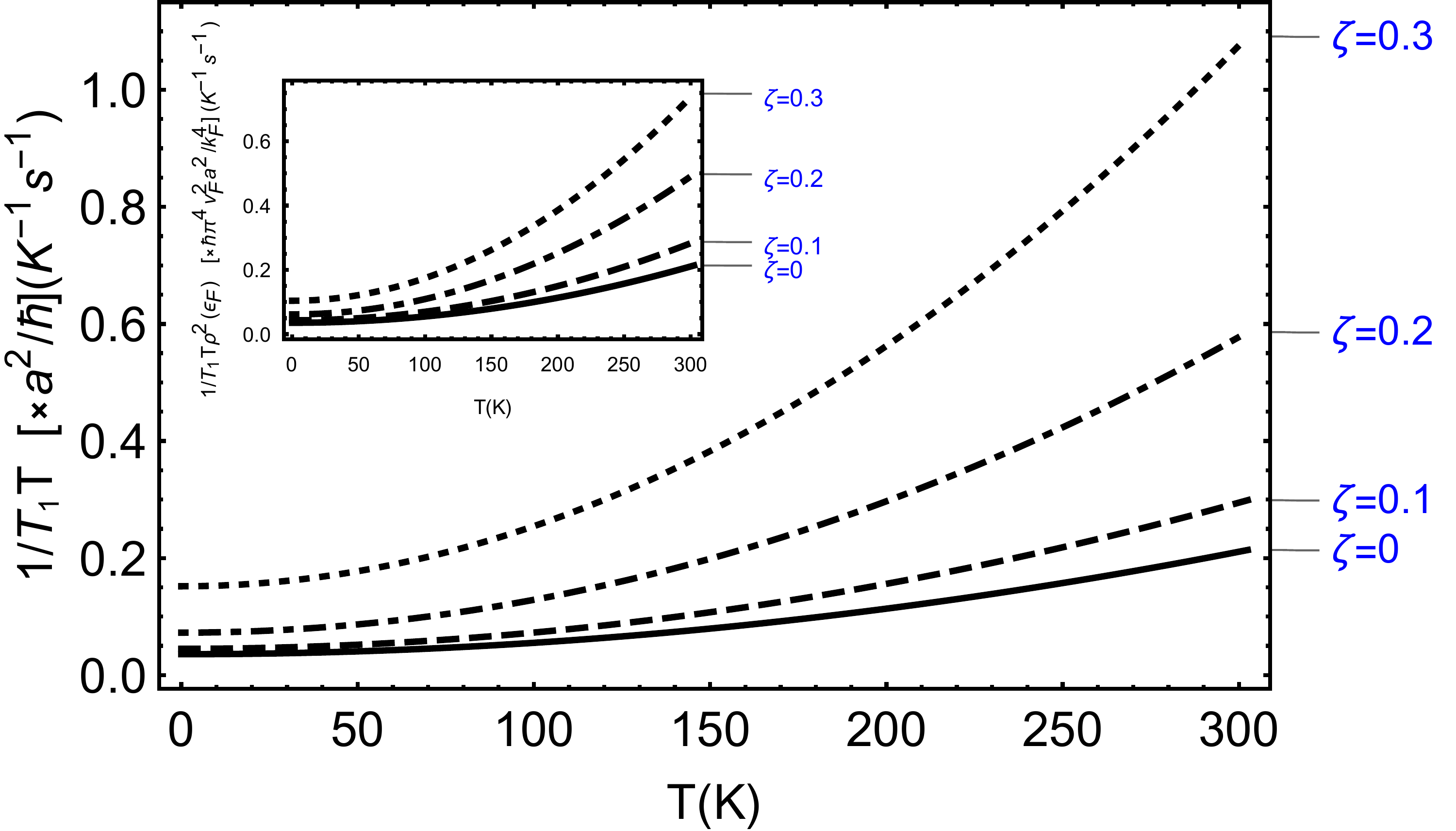} 
   \caption{The temperature dependence of nuclear spin relaxation rate $1/{T_1 T}$ for fixed values of tilt 
parameters $\zeta$. The quantity $a$ is defined by $a=\hbar v_F \mu_0\gamma_0 e$ where $v_F\sim 10^6 m/s$ is typical Fermi velocity of quasiparticles and $k_F\sim 1.5 \times 10^8 m^{-1}$ is Fermi wavevector. The inset excludes the contribution of DOS at Fermi level. 
}
\label{Fig1}
\end{figure}

In $k_B T \ll \mu $ regime, the Sommerfeld expansion reduces the integral in Eq.~\eqref{eq:10} to integration over solid angles $\Omega_{\bs k}$ and $\Omega_{\bs k^\prime}$. 
For $k_B T \ll \mu $ and $\zeta=0$ (non-tilted case) NMR rate has a weak dependence on temperature indicated in Fig.\ref{Fig1} 
which is the conventional behavior (black solid line). Indeed within the Korringa relation the $1/(T_1T)$ is independent of $T$. This corresponds to $T\sim 0$ in
Fig.~\ref{Fig1}. The additional parabolic dependence on $T$ comes from our Sommerfeld expansion upto this order. 
Upon introducing the tilt ($\zeta\neq 0$) spin lattice relaxation rate acquires notably sensitive to temperature. This sensitivity is further enhanced upon
further increasing the tilt parameter. One might argue that the enhancement of the NMR relaxation rate in TCWSMs is a DOS effect. 
This seems feasible, bacause according to Korringa relation, the relaxation rate is proportional to the square of the density of states. 
To investigate this, in the inset of Fig.\ref{Fig1} we have excluded the $\rho^2$ in order to separate the sole effect of the tilt parameter. As can be seen even after 
excluding the effect of DOS, while preserving the parabolic temperature dependence, still the relaxation rate shows enhancement caused by 
tilt parameter $\zeta$. Even the low-T part of this figure that corresponds to Korringa limit, shows tilt induced enhancement.

\begin{figure}[b]
\includegraphics[width=0.5\textwidth]{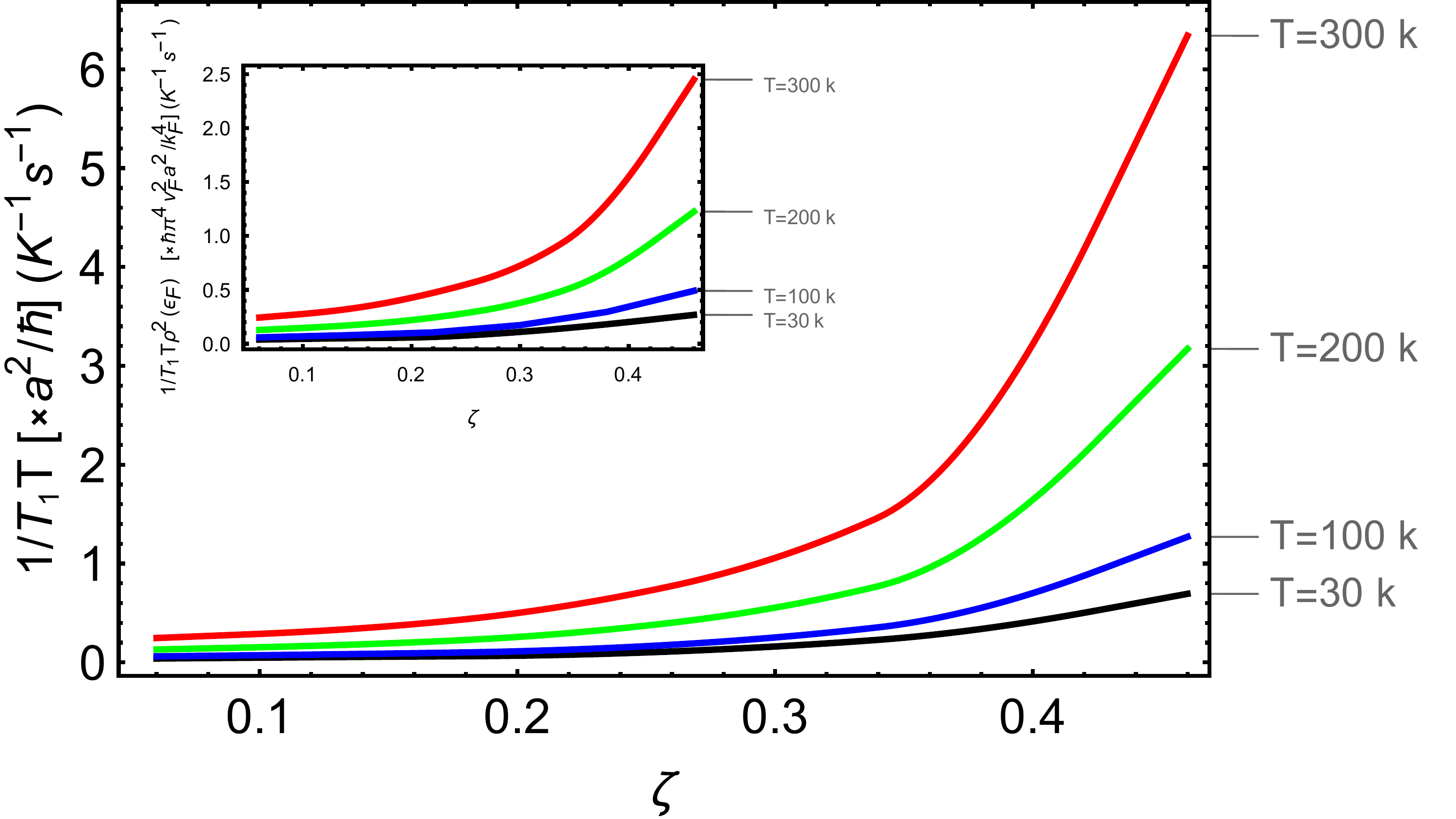}   
\caption{The nuclear spin relaxation rate $1/{T_1 T}$ as a function of tilt parameter $\zeta$ at fixed temperatures. The quantity $a$ is defined by $a=\hbar v_F \mu_0\gamma_0 e$ where $v_F\sim 10^6 m/s$ is typical Fermi velocity of quasiparticles and $k_F\sim 1.5 \times 10^8 m^{-1}$ is Fermi wavevector. The inset is the same, but excluding the DOS effect.}
\label{Fig2}
\end{figure}

To further investigate the role of the tilt parameter in relaxation time, we split it into two types of enhancements. 
First type is the enhancement of the DOS by the tilt paramter which as discussed below Eq.~\eqref{eq:2} is proportional to $(1-\zeta^2)^{-2}$.
Within a Korringa formula, this will give rise to a $\rho^2\sim (1-\zeta^2)^{-4}$ factor. 
There is still a significant dependence on $\zeta$ that comes through the matrix elements of the hyperfine interaction. 
Fig.~\ref{Fig2} illustrates that $(T_1T)^{-1}$ depends on $\zeta$. The inset is the same, excluding a $\rho^2$ factor.
As can be seen, even after excluding the $\rho^2$ factor, still a divergence at $\zeta\to 1$ persists. 
It appears that the tilt parameter is acting as a new relaxation channel that paves the way for nuclear spins to relax back. 
Fitting a $(1-\zeta^2)^{-\alpha}$ behavior gives a value of $\alpha_0=8.927\pm 0.024$ for $T=0$ when the $90$ data points in the range $0.9\le \zeta \le 0.99$
are used for the fit. For $T=300K$ the exponent $\alpha=9.07$. 

What is the source of such a strong divergence with $\alpha=9$ in the limit $\zeta\to 1$? To investigate this, note that in Eq.~\eqref{eq:10}, 
there are two Jacobians containing $(1+\zeta\cos\theta)^3$ factors in the denominator. If one deliberately drops these factors
from the integral, the remaning integral, instead of diverging with $\alpha= 9$, the result will diverge with $\alpha_{\rm hf}=3.927\pm 0.025\approx 4$ at zero temperature,
where the subscript "hf" is used to emphasize the part of divergence arising from the hyperfine matrix elements without including the Jacobian. 
Therefore the product of two Jacobian contributes an exponent $\alpha_{\rm J}=5$ leaving an exponent of $2.5$ for each Jacobian factors. 
Therefore a strong divergence with an exponent of $\alpha=9$ is composed of Jacobian part $\alpha_{\rm J}=5$ as well as hyperfine contributions $\alpha_{\rm hf}=4$.

\section{discussioins and summary}
In this paper we studied the NMR relaxation rate for a tilted cone Weyl semimetal. We found that 
the contribution of tilt parameter is more than an additional coupling term in hyperfine interaction. Since in TCWSMs we have ellipsoidal Fermi surface
whose eccentricity turns out to be given by the magnitude $\zeta$ of the tilt parameter, the spin and orbital parts immediately imprint their $\zeta$ dependence 
into the NMR relaxation rate as a very fast relaxation rate that diverges like $(1-\zeta^2)^{-\alpha}$ with $\alpha\approx 9$ for $\zeta\lesssim 1$. 
The interpretation of the fast relaxation is that the tilt provides an additional relaxation channel for the
nuclear spins via additional coupling of the spin and orbital hyperfine interaction on tilt parameter that accelerates the relaxation process. 

In order to understand the exponent $\alpha\approx 9$ in the NMR relaxation rate, let us start by assuming that the tilt in the dispersion of TCWSMs that mixes energy-momentum is rooted in the metric 
$ds^2=-v_F^2dt^2+(d{\bs r}-\bs\zeta v_Fdt)^2$~\cite{Tohid2019Spacetime,Sahar2019Polarization}. Let us further assume 
that in some part of this spacetime we have a given non-zero $\zeta$ and in some other corner we have $\zeta=0$. 
The time intervals in these two parts of the spacetime are $dt$ and $dt_0$, respectively which are related by 
$dt=dt_0/\sqrt{1-\zeta^2}$ where $\sqrt{1-\zeta^2}$ is the "gravitational" redshift factor. Now let us turn our
attention to the NMR rates in TCWSMs. The power $4$ out of $9$ was shown to arise from non-separable hyperfine matrix elements
when they are all integrated together. Within the Korringa framework, insisting that the remaining power of $5$ arises from
a DOS squared, $\tilde \rho^2$, one concludes that $\tilde\rho\sim (1-\zeta^2)^{-2.5}$. But on the other hand, the 
DOS calculated from the tilted conic dispersion is $\rho\sim (1-\zeta^2)^{-2}$. We therefore find that $\tilde\rho=\rho/\sqrt{1-\zeta^2}$. 

In the present NMR relaxation theory, the quantity $1/(T_1T)$ contains the product of "time" and "energy" scales, and therefore one does not
expect to directly observe the redshift factor in such quantity. However in one hand the density of states has the dimension of 
inverse energy (i.e. the dimension of "time" in units with $\hbar=1$) which makes it a suitable quantity to look for redshift 
factor in TCWSMs. On the other hand, the local nature of NMR experiments allows to compare the density $\tilde\rho$ of states inferred from
Korringa relation as a function of tilt parameter $\zeta$ with the bare DOS $\rho$ obtained from band structure at every point with 
a given $\zeta$. Then these two "time scales" are related by "gravitational" redshift factor. 

\section{Acknowledgements}
S.A.J. was supported by research deputy of Sharif University of Technology, grant no. G960214 and the
Iran Science Elites Federation (ISEF). Z.F. was supported by ISEF post doctoral fellowship.

\appendix
\section{}
\numberwithin{equation}{section}
\makeatletter 
\newcommand{\section@cntformat}{appendix \thesection:\ }
\makeatother
\setcounter{equation}{0}
To obtain a proper coordinates for our system, we start with upper band dispersion for $\boldsymbol{\zeta}=\zeta\ \hat{\textbf{z}}$, 
that reads $ \epsilon=\zeta\ k_z +|\bs{k}|$, or alternatively, $|\bs{k}|={\varepsilon}/{(1+\zeta\cos\theta)}$ which gives
the Cartesian components of the momentum as
\be 
   (k_x,k_y,k_z)=\frac{\varepsilon (\sin\theta\cos\phi,\sin\theta\sin\phi,\cos\theta) }{1+\zeta\cos\theta} 
\ee
that provides a natural generalization of the spherical cordinate with $\theta$ and $\phi$ being polar and
azimuthal angles. The Jacobian of the new coordinates transformation will be
$
  J(\varepsilon,\theta,\phi)={\varepsilon^2 \sin\theta}/{(1+\zeta \cos \theta)^3},
$
Integrating over the Fermi surface with a constant $\varepsilon=\varepsilon_F$ gives rise to the density of states 
and the $(1-\zeta^2)^{-2}$ enhancement. On the other hand by standard contour integration one can show that the strongest
divergence in $\zeta\to 1$ that can be contributed by a factor like 
$(1+\zeta\cos\theta)^{-n}$ is given by $(1-\zeta^2)^{0.5-n}$.

\bibliography{mybib}

\begin{thebibliography}{40}%
\makeatletter
\providecommand \@ifxundefined [1]{%
 \@ifx{#1\undefined}
}%
\providecommand \@ifnum [1]{%
 \ifnum #1\expandafter \@firstoftwo
 \else \expandafter \@secondoftwo
 \fi
}%
\providecommand \@ifx [1]{%
 \ifx #1\expandafter \@firstoftwo
 \else \expandafter \@secondoftwo
 \fi
}%
\providecommand \natexlab [1]{#1}%
\providecommand \enquote  [1]{``#1''}%
\providecommand \bibnamefont  [1]{#1}%
\providecommand \bibfnamefont [1]{#1}%
\providecommand \citenamefont [1]{#1}%
\providecommand \href@noop [0]{\@secondoftwo}%
\providecommand \href [0]{\begingroup \@sanitize@url \@href}%
\providecommand \@href[1]{\@@startlink{#1}\@@href}%
\providecommand \@@href[1]{\endgroup#1\@@endlink}%
\providecommand \@sanitize@url [0]{\catcode `\\12\catcode `\$12\catcode
  `\&12\catcode `\#12\catcode `\^12\catcode `\_12\catcode `\%12\relax}%
\providecommand \@@startlink[1]{}%
\providecommand \@@endlink[0]{}%
\providecommand \url  [0]{\begingroup\@sanitize@url \@url }%
\providecommand \@url [1]{\endgroup\@href {#1}{\urlprefix }}%
\providecommand \urlprefix  [0]{URL }%
\providecommand \Eprint [0]{\href }%
\providecommand \doibase [0]{http://dx.doi.org/}%
\providecommand \selectlanguage [0]{\@gobble}%
\providecommand \bibinfo  [0]{\@secondoftwo}%
\providecommand \bibfield  [0]{\@secondoftwo}%
\providecommand \translation [1]{[#1]}%
\providecommand \BibitemOpen [0]{}%
\providecommand \bibitemStop [0]{}%
\providecommand \bibitemNoStop [0]{.\EOS\space}%
\providecommand \EOS [0]{\spacefactor3000\relax}%
\providecommand \BibitemShut  [1]{\csname bibitem#1\endcsname}%
\let\auto@bib@innerbib\@empty
\bibitem [{\citenamefont {Xu}\ \emph {et~al.}(2015{\natexlab{a}})\citenamefont
  {Xu}, \citenamefont {Belopolski}, \citenamefont {Alidoust}, \citenamefont
  {Neupane}, \citenamefont {Bian}, \citenamefont {Zhang}, \citenamefont
  {Sankar}, \citenamefont {Chang}, \citenamefont {Yuan}, \citenamefont {Lee}
  \emph {et~al.}}]{xu2015discovery}%
  \BibitemOpen
  \bibfield  {author} {\bibinfo {author} {\bibfnamefont {S.-Y.}\ \bibnamefont
  {Xu}}, \bibinfo {author} {\bibfnamefont {I.}~\bibnamefont {Belopolski}},
  \bibinfo {author} {\bibfnamefont {N.}~\bibnamefont {Alidoust}}, \bibinfo
  {author} {\bibfnamefont {M.}~\bibnamefont {Neupane}}, \bibinfo {author}
  {\bibfnamefont {G.}~\bibnamefont {Bian}}, \bibinfo {author} {\bibfnamefont
  {C.}~\bibnamefont {Zhang}}, \bibinfo {author} {\bibfnamefont
  {R.}~\bibnamefont {Sankar}}, \bibinfo {author} {\bibfnamefont
  {G.}~\bibnamefont {Chang}}, \bibinfo {author} {\bibfnamefont
  {Z.}~\bibnamefont {Yuan}}, \bibinfo {author} {\bibfnamefont {C.-C.}\
  \bibnamefont {Lee}},  \emph {et~al.},\ }\href
  {https://science.sciencemag.org/content/349/6248/613} {\bibfield  {journal}
  {\bibinfo  {journal} {Science}\ }\textbf {\bibinfo {volume} {349}},\ \bibinfo
  {pages} {613} (\bibinfo {year} {2015}{\natexlab{a}})}\BibitemShut {NoStop}%
\bibitem [{\citenamefont {Lv}\ \emph {et~al.}(2015)\citenamefont {Lv},
  \citenamefont {Weng}, \citenamefont {Fu}, \citenamefont {Wang}, \citenamefont
  {Miao}, \citenamefont {Ma}, \citenamefont {Richard}, \citenamefont {Huang},
  \citenamefont {Zhao}, \citenamefont {Chen} \emph
  {et~al.}}]{lv2015experimental}%
  \BibitemOpen
  \bibfield  {author} {\bibinfo {author} {\bibfnamefont {B.}~\bibnamefont
  {Lv}}, \bibinfo {author} {\bibfnamefont {H.}~\bibnamefont {Weng}}, \bibinfo
  {author} {\bibfnamefont {B.}~\bibnamefont {Fu}}, \bibinfo {author}
  {\bibfnamefont {X.}~\bibnamefont {Wang}}, \bibinfo {author} {\bibfnamefont
  {H.}~\bibnamefont {Miao}}, \bibinfo {author} {\bibfnamefont {J.}~\bibnamefont
  {Ma}}, \bibinfo {author} {\bibfnamefont {P.}~\bibnamefont {Richard}},
  \bibinfo {author} {\bibfnamefont {X.}~\bibnamefont {Huang}}, \bibinfo
  {author} {\bibfnamefont {L.}~\bibnamefont {Zhao}}, \bibinfo {author}
  {\bibfnamefont {G.}~\bibnamefont {Chen}},  \emph {et~al.},\ }\href
  {https://journals.aps.org/prx/abstract/10.1103/PhysRevX.5.031013} {\bibfield
  {journal} {\bibinfo  {journal} {Physical Review X}\ }\textbf {\bibinfo
  {volume} {5}},\ \bibinfo {pages} {031013} (\bibinfo {year}
  {2015})}\BibitemShut {NoStop}%
\bibitem [{\citenamefont {Yang}\ \emph
  {et~al.}(2015{\natexlab{a}})\citenamefont {Yang}, \citenamefont {Liu},
  \citenamefont {Sun}, \citenamefont {Peng}, \citenamefont {Yang},
  \citenamefont {Zhang}, \citenamefont {Zhou}, \citenamefont {Zhang},
  \citenamefont {Guo}, \citenamefont {Rahn} \emph {et~al.}}]{yang2015weyl}%
  \BibitemOpen
  \bibfield  {author} {\bibinfo {author} {\bibfnamefont {L.}~\bibnamefont
  {Yang}}, \bibinfo {author} {\bibfnamefont {Z.}~\bibnamefont {Liu}}, \bibinfo
  {author} {\bibfnamefont {Y.}~\bibnamefont {Sun}}, \bibinfo {author}
  {\bibfnamefont {H.}~\bibnamefont {Peng}}, \bibinfo {author} {\bibfnamefont
  {H.}~\bibnamefont {Yang}}, \bibinfo {author} {\bibfnamefont {T.}~\bibnamefont
  {Zhang}}, \bibinfo {author} {\bibfnamefont {B.}~\bibnamefont {Zhou}},
  \bibinfo {author} {\bibfnamefont {Y.}~\bibnamefont {Zhang}}, \bibinfo
  {author} {\bibfnamefont {Y.}~\bibnamefont {Guo}}, \bibinfo {author}
  {\bibfnamefont {M.}~\bibnamefont {Rahn}},  \emph {et~al.},\ }\href
  {https://www.nature.com/articles/nphys3425} {\bibfield  {journal} {\bibinfo
  {journal} {Nature physics}\ }\textbf {\bibinfo {volume} {11}},\ \bibinfo
  {pages} {728} (\bibinfo {year} {2015}{\natexlab{a}})}\BibitemShut {NoStop}%
\bibitem [{\citenamefont {Zheng}\ \emph {et~al.}(2016)\citenamefont {Zheng},
  \citenamefont {Xu}, \citenamefont {Bian}, \citenamefont {Guo}, \citenamefont
  {Chang}, \citenamefont {Sanchez}, \citenamefont {Belopolski}, \citenamefont
  {Lee}, \citenamefont {Huang}, \citenamefont {Zhang} \emph
  {et~al.}}]{zheng2016atomic}%
  \BibitemOpen
  \bibfield  {author} {\bibinfo {author} {\bibfnamefont {H.}~\bibnamefont
  {Zheng}}, \bibinfo {author} {\bibfnamefont {S.-Y.}\ \bibnamefont {Xu}},
  \bibinfo {author} {\bibfnamefont {G.}~\bibnamefont {Bian}}, \bibinfo {author}
  {\bibfnamefont {C.}~\bibnamefont {Guo}}, \bibinfo {author} {\bibfnamefont
  {G.}~\bibnamefont {Chang}}, \bibinfo {author} {\bibfnamefont {D.~S.}\
  \bibnamefont {Sanchez}}, \bibinfo {author} {\bibfnamefont {I.}~\bibnamefont
  {Belopolski}}, \bibinfo {author} {\bibfnamefont {C.-C.}\ \bibnamefont {Lee}},
  \bibinfo {author} {\bibfnamefont {S.-M.}\ \bibnamefont {Huang}}, \bibinfo
  {author} {\bibfnamefont {X.}~\bibnamefont {Zhang}},  \emph {et~al.},\ }\href
  {https://pubs.acs.org/doi/abs/10.1021/acsnano.5b06807} {\bibfield  {journal}
  {\bibinfo  {journal} {ACS nano}\ }\textbf {\bibinfo {volume} {10}},\ \bibinfo
  {pages} {1378} (\bibinfo {year} {2016})}\BibitemShut {NoStop}%
\bibitem [{\citenamefont {Liu}\ \emph {et~al.}(2016)\citenamefont {Liu},
  \citenamefont {Yang}, \citenamefont {Sun}, \citenamefont {Zhang},
  \citenamefont {Peng}, \citenamefont {Yang}, \citenamefont {Chen},
  \citenamefont {Zhang}, \citenamefont {Guo}, \citenamefont {Prabhakaran} \emph
  {et~al.}}]{liu2016evolution}%
  \BibitemOpen
  \bibfield  {author} {\bibinfo {author} {\bibfnamefont {Z.}~\bibnamefont
  {Liu}}, \bibinfo {author} {\bibfnamefont {L.}~\bibnamefont {Yang}}, \bibinfo
  {author} {\bibfnamefont {Y.}~\bibnamefont {Sun}}, \bibinfo {author}
  {\bibfnamefont {T.}~\bibnamefont {Zhang}}, \bibinfo {author} {\bibfnamefont
  {H.}~\bibnamefont {Peng}}, \bibinfo {author} {\bibfnamefont {H.}~\bibnamefont
  {Yang}}, \bibinfo {author} {\bibfnamefont {C.}~\bibnamefont {Chen}}, \bibinfo
  {author} {\bibfnamefont {Y.}~\bibnamefont {Zhang}}, \bibinfo {author}
  {\bibfnamefont {Y.}~\bibnamefont {Guo}}, \bibinfo {author} {\bibfnamefont
  {D.}~\bibnamefont {Prabhakaran}},  \emph {et~al.},\ }\href
  {https://www.nature.com/articles/nmat4457} {\bibfield  {journal} {\bibinfo
  {journal} {Nature materials}\ }\textbf {\bibinfo {volume} {15}},\ \bibinfo
  {pages} {27} (\bibinfo {year} {2016})}\BibitemShut {NoStop}%
\bibitem [{\citenamefont {Xu}\ \emph {et~al.}(2015{\natexlab{b}})\citenamefont
  {Xu}, \citenamefont {Belopolski}, \citenamefont {Sanchez}, \citenamefont
  {Zhang}, \citenamefont {Chang}, \citenamefont {Guo}, \citenamefont {Bian},
  \citenamefont {Yuan}, \citenamefont {Lu}, \citenamefont {Chang} \emph
  {et~al.}}]{xu2015experimental}%
  \BibitemOpen
  \bibfield  {author} {\bibinfo {author} {\bibfnamefont {S.-Y.}\ \bibnamefont
  {Xu}}, \bibinfo {author} {\bibfnamefont {I.}~\bibnamefont {Belopolski}},
  \bibinfo {author} {\bibfnamefont {D.~S.}\ \bibnamefont {Sanchez}}, \bibinfo
  {author} {\bibfnamefont {C.}~\bibnamefont {Zhang}}, \bibinfo {author}
  {\bibfnamefont {G.}~\bibnamefont {Chang}}, \bibinfo {author} {\bibfnamefont
  {C.}~\bibnamefont {Guo}}, \bibinfo {author} {\bibfnamefont {G.}~\bibnamefont
  {Bian}}, \bibinfo {author} {\bibfnamefont {Z.}~\bibnamefont {Yuan}}, \bibinfo
  {author} {\bibfnamefont {H.}~\bibnamefont {Lu}}, \bibinfo {author}
  {\bibfnamefont {T.-R.}\ \bibnamefont {Chang}},  \emph {et~al.},\ }\href
  {https://advances.sciencemag.org/content/1/10/e1501092?utm_source=TrendMD&utm_medium=cpc&utm_campaign=TrendMD_1}
  {\bibfield  {journal} {\bibinfo  {journal} {Science advances}\ }\textbf
  {\bibinfo {volume} {1}},\ \bibinfo {pages} {e1501092} (\bibinfo {year}
  {2015}{\natexlab{b}})}\BibitemShut {NoStop}%
\bibitem [{\citenamefont {Xu}\ \emph {et~al.}(2016)\citenamefont {Xu},
  \citenamefont {Weng}, \citenamefont {Lv}, \citenamefont {Matt}, \citenamefont
  {Park}, \citenamefont {Bisti}, \citenamefont {Strocov}, \citenamefont
  {Gawryluk}, \citenamefont {Pomjakushina}, \citenamefont {Conder} \emph
  {et~al.}}]{xu2016observation}%
  \BibitemOpen
  \bibfield  {author} {\bibinfo {author} {\bibfnamefont {N.}~\bibnamefont
  {Xu}}, \bibinfo {author} {\bibfnamefont {H.}~\bibnamefont {Weng}}, \bibinfo
  {author} {\bibfnamefont {B.}~\bibnamefont {Lv}}, \bibinfo {author}
  {\bibfnamefont {C.~E.}\ \bibnamefont {Matt}}, \bibinfo {author}
  {\bibfnamefont {J.}~\bibnamefont {Park}}, \bibinfo {author} {\bibfnamefont
  {F.}~\bibnamefont {Bisti}}, \bibinfo {author} {\bibfnamefont {V.~N.}\
  \bibnamefont {Strocov}}, \bibinfo {author} {\bibfnamefont {D.}~\bibnamefont
  {Gawryluk}}, \bibinfo {author} {\bibfnamefont {E.}~\bibnamefont
  {Pomjakushina}}, \bibinfo {author} {\bibfnamefont {K.}~\bibnamefont
  {Conder}},  \emph {et~al.},\ }\href
  {https://www.nature.com/articles/ncomms11006} {\bibfield  {journal} {\bibinfo
   {journal} {Nature communications}\ }\textbf {\bibinfo {volume} {7}},\
  \bibinfo {pages} {1} (\bibinfo {year} {2016})}\BibitemShut {NoStop}%
\bibitem [{\citenamefont {Belopolski}\ \emph {et~al.}(2016)\citenamefont
  {Belopolski}, \citenamefont {Xu}, \citenamefont {Sanchez}, \citenamefont
  {Chang}, \citenamefont {Guo}, \citenamefont {Neupane}, \citenamefont {Zheng},
  \citenamefont {Lee}, \citenamefont {Huang}, \citenamefont {Bian} \emph
  {et~al.}}]{belopolski2016criteria}%
  \BibitemOpen
  \bibfield  {author} {\bibinfo {author} {\bibfnamefont {I.}~\bibnamefont
  {Belopolski}}, \bibinfo {author} {\bibfnamefont {S.-Y.}\ \bibnamefont {Xu}},
  \bibinfo {author} {\bibfnamefont {D.~S.}\ \bibnamefont {Sanchez}}, \bibinfo
  {author} {\bibfnamefont {G.}~\bibnamefont {Chang}}, \bibinfo {author}
  {\bibfnamefont {C.}~\bibnamefont {Guo}}, \bibinfo {author} {\bibfnamefont
  {M.}~\bibnamefont {Neupane}}, \bibinfo {author} {\bibfnamefont
  {H.}~\bibnamefont {Zheng}}, \bibinfo {author} {\bibfnamefont {C.-C.}\
  \bibnamefont {Lee}}, \bibinfo {author} {\bibfnamefont {S.-M.}\ \bibnamefont
  {Huang}}, \bibinfo {author} {\bibfnamefont {G.}~\bibnamefont {Bian}},  \emph
  {et~al.},\ }\href
  {https://journals.aps.org/prl/abstract/10.1103/PhysRevLett.116.066802}
  {\bibfield  {journal} {\bibinfo  {journal} {Physical review letters}\
  }\textbf {\bibinfo {volume} {116}},\ \bibinfo {pages} {066802} (\bibinfo
  {year} {2016})}\BibitemShut {NoStop}%
\bibitem [{\citenamefont {Souma}\ \emph {et~al.}(2016)\citenamefont {Souma},
  \citenamefont {Wang}, \citenamefont {Kotaka}, \citenamefont {Sato},
  \citenamefont {Nakayama}, \citenamefont {Tanaka}, \citenamefont {Kimizuka},
  \citenamefont {Takahashi}, \citenamefont {Yamauchi}, \citenamefont {Oguchi}
  \emph {et~al.}}]{souma2016direct}%
  \BibitemOpen
  \bibfield  {author} {\bibinfo {author} {\bibfnamefont {S.}~\bibnamefont
  {Souma}}, \bibinfo {author} {\bibfnamefont {Z.}~\bibnamefont {Wang}},
  \bibinfo {author} {\bibfnamefont {H.}~\bibnamefont {Kotaka}}, \bibinfo
  {author} {\bibfnamefont {T.}~\bibnamefont {Sato}}, \bibinfo {author}
  {\bibfnamefont {K.}~\bibnamefont {Nakayama}}, \bibinfo {author}
  {\bibfnamefont {Y.}~\bibnamefont {Tanaka}}, \bibinfo {author} {\bibfnamefont
  {H.}~\bibnamefont {Kimizuka}}, \bibinfo {author} {\bibfnamefont
  {T.}~\bibnamefont {Takahashi}}, \bibinfo {author} {\bibfnamefont
  {K.}~\bibnamefont {Yamauchi}}, \bibinfo {author} {\bibfnamefont
  {T.}~\bibnamefont {Oguchi}},  \emph {et~al.},\ }\href
  {https://journals.aps.org/prb/abstract/10.1103/PhysRevB.93.161112} {\bibfield
   {journal} {\bibinfo  {journal} {Physical Review B}\ }\textbf {\bibinfo
  {volume} {93}},\ \bibinfo {pages} {161112} (\bibinfo {year}
  {2016})}\BibitemShut {NoStop}%
\bibitem [{\citenamefont {Xu}\ \emph {et~al.}(2017)\citenamefont {Xu},
  \citenamefont {Autes}, \citenamefont {Matt}, \citenamefont {Lv},
  \citenamefont {Yao}, \citenamefont {Bisti}, \citenamefont {Strocov},
  \citenamefont {Gawryluk}, \citenamefont {Pomjakushina}, \citenamefont
  {Conder} \emph {et~al.}}]{xu2017distinct}%
  \BibitemOpen
  \bibfield  {author} {\bibinfo {author} {\bibfnamefont {N.}~\bibnamefont
  {Xu}}, \bibinfo {author} {\bibfnamefont {G.}~\bibnamefont {Autes}}, \bibinfo
  {author} {\bibfnamefont {C.~E.}\ \bibnamefont {Matt}}, \bibinfo {author}
  {\bibfnamefont {B.}~\bibnamefont {Lv}}, \bibinfo {author} {\bibfnamefont
  {M.}~\bibnamefont {Yao}}, \bibinfo {author} {\bibfnamefont {F.}~\bibnamefont
  {Bisti}}, \bibinfo {author} {\bibfnamefont {V.~N.}\ \bibnamefont {Strocov}},
  \bibinfo {author} {\bibfnamefont {D.}~\bibnamefont {Gawryluk}}, \bibinfo
  {author} {\bibfnamefont {E.}~\bibnamefont {Pomjakushina}}, \bibinfo {author}
  {\bibfnamefont {K.}~\bibnamefont {Conder}},  \emph {et~al.},\ }\href
  {https://journals.aps.org/prl/abstract/10.1103/PhysRevLett.118.106406}
  {\bibfield  {journal} {\bibinfo  {journal} {Physical review letters}\
  }\textbf {\bibinfo {volume} {118}},\ \bibinfo {pages} {106406} (\bibinfo
  {year} {2017})}\BibitemShut {NoStop}%
\bibitem [{\citenamefont {Borisenko}\ \emph {et~al.}(2019)\citenamefont
  {Borisenko}, \citenamefont {Evtushinsky}, \citenamefont {Gibson},
  \citenamefont {Yaresko}, \citenamefont {Koepernik}, \citenamefont {Kim},
  \citenamefont {Ali}, \citenamefont {van~den Brink}, \citenamefont {Hoesch},
  \citenamefont {Fedorov} \emph {et~al.}}]{borisenko2019time}%
  \BibitemOpen
  \bibfield  {author} {\bibinfo {author} {\bibfnamefont {S.}~\bibnamefont
  {Borisenko}}, \bibinfo {author} {\bibfnamefont {D.}~\bibnamefont
  {Evtushinsky}}, \bibinfo {author} {\bibfnamefont {Q.}~\bibnamefont {Gibson}},
  \bibinfo {author} {\bibfnamefont {A.}~\bibnamefont {Yaresko}}, \bibinfo
  {author} {\bibfnamefont {K.}~\bibnamefont {Koepernik}}, \bibinfo {author}
  {\bibfnamefont {T.}~\bibnamefont {Kim}}, \bibinfo {author} {\bibfnamefont
  {M.}~\bibnamefont {Ali}}, \bibinfo {author} {\bibfnamefont {J.}~\bibnamefont
  {van~den Brink}}, \bibinfo {author} {\bibfnamefont {M.}~\bibnamefont
  {Hoesch}}, \bibinfo {author} {\bibfnamefont {A.}~\bibnamefont {Fedorov}},
  \emph {et~al.},\ }\href {https://www.nature.com/articles/s41467-019-11393-5}
  {\bibfield  {journal} {\bibinfo  {journal} {Nature communications}\ }\textbf
  {\bibinfo {volume} {10}},\ \bibinfo {pages} {1} (\bibinfo {year}
  {2019})}\BibitemShut {NoStop}%
\bibitem [{\citenamefont {Weyl}(1929)}]{weyl1929elektron}%
  \BibitemOpen
  \bibfield  {author} {\bibinfo {author} {\bibfnamefont {H.}~\bibnamefont
  {Weyl}},\ }\href {https://link.springer.com/article/10.1007/BF01339504}
  {\bibfield  {journal} {\bibinfo  {journal} {Zeitschrift f{\"u}r Physik}\
  }\textbf {\bibinfo {volume} {56}},\ \bibinfo {pages} {330} (\bibinfo {year}
  {1929})}\BibitemShut {NoStop}%
\bibitem [{\citenamefont {Zheng}\ and\ \citenamefont
  {Zahid~Hasan}(2018)}]{zheng2018quasiparticle}%
  \BibitemOpen
  \bibfield  {author} {\bibinfo {author} {\bibfnamefont {H.}~\bibnamefont
  {Zheng}}\ and\ \bibinfo {author} {\bibfnamefont {M.}~\bibnamefont
  {Zahid~Hasan}},\ }\href
  {https://www.tandfonline.com/doi/full/10.1080/23746149.2018.1466661}
  {\bibfield  {journal} {\bibinfo  {journal} {Advances in Physics: X}\ }\textbf
  {\bibinfo {volume} {3}},\ \bibinfo {pages} {1466661} (\bibinfo {year}
  {2018})}\BibitemShut {NoStop}%
\bibitem [{\citenamefont {Wehling}\ \emph {et~al.}(2014)\citenamefont
  {Wehling}, \citenamefont {Black-Schaffer},\ and\ \citenamefont
  {Balatsky}}]{wehling2014dirac}%
  \BibitemOpen
  \bibfield  {author} {\bibinfo {author} {\bibfnamefont {T.}~\bibnamefont
  {Wehling}}, \bibinfo {author} {\bibfnamefont {A.~M.}\ \bibnamefont
  {Black-Schaffer}}, \ and\ \bibinfo {author} {\bibfnamefont {A.~V.}\
  \bibnamefont {Balatsky}},\ }\href
  {https://www.tandfonline.com/doi/abs/10.1080/00018732.2014.927109} {\bibfield
   {journal} {\bibinfo  {journal} {Advances in Physics}\ }\textbf {\bibinfo
  {volume} {63}},\ \bibinfo {pages} {1} (\bibinfo {year} {2014})}\BibitemShut
  {NoStop}%
\bibitem [{\citenamefont {Faraei}\ and\ \citenamefont
  {Jafari}(2017)}]{faraei2017superconducting}%
  \BibitemOpen
  \bibfield  {author} {\bibinfo {author} {\bibfnamefont {Z.}~\bibnamefont
  {Faraei}}\ and\ \bibinfo {author} {\bibfnamefont {S.}~\bibnamefont
  {Jafari}},\ }\href
  {https://journals.aps.org/prb/abstract/10.1103/PhysRevB.96.134516} {\bibfield
   {journal} {\bibinfo  {journal} {Physical Review B}\ }\textbf {\bibinfo
  {volume} {96}},\ \bibinfo {pages} {134516} (\bibinfo {year}
  {2017})}\BibitemShut {NoStop}%
\bibitem [{\citenamefont {Son}\ and\ \citenamefont
  {Spivak}(2013)}]{son2013chiral}%
  \BibitemOpen
  \bibfield  {author} {\bibinfo {author} {\bibfnamefont {D.}~\bibnamefont
  {Son}}\ and\ \bibinfo {author} {\bibfnamefont {B.}~\bibnamefont {Spivak}},\
  }\href {https://journals.aps.org/prb/abstract/10.1103/PhysRevB.88.104412}
  {\bibfield  {journal} {\bibinfo  {journal} {Physical Review B}\ }\textbf
  {\bibinfo {volume} {88}},\ \bibinfo {pages} {104412} (\bibinfo {year}
  {2013})}\BibitemShut {NoStop}%
\bibitem [{\citenamefont {Parameswaran}\ \emph {et~al.}(2014)\citenamefont
  {Parameswaran}, \citenamefont {Grover}, \citenamefont {Abanin}, \citenamefont
  {Pesin},\ and\ \citenamefont {Vishwanath}}]{parameswaran2014probing}%
  \BibitemOpen
  \bibfield  {author} {\bibinfo {author} {\bibfnamefont {S.}~\bibnamefont
  {Parameswaran}}, \bibinfo {author} {\bibfnamefont {T.}~\bibnamefont
  {Grover}}, \bibinfo {author} {\bibfnamefont {D.}~\bibnamefont {Abanin}},
  \bibinfo {author} {\bibfnamefont {D.}~\bibnamefont {Pesin}}, \ and\ \bibinfo
  {author} {\bibfnamefont {A.}~\bibnamefont {Vishwanath}},\ }\href
  {https://journals.aps.org/prx/abstract/10.1103/PhysRevX.4.031035} {\bibfield
  {journal} {\bibinfo  {journal} {Physical Review X}\ }\textbf {\bibinfo
  {volume} {4}},\ \bibinfo {pages} {031035} (\bibinfo {year}
  {2014})}\BibitemShut {NoStop}%
\bibitem [{\citenamefont {Yang}\ \emph {et~al.}(2011)\citenamefont {Yang},
  \citenamefont {Lu},\ and\ \citenamefont {Ran}}]{yang2011quantum}%
  \BibitemOpen
  \bibfield  {author} {\bibinfo {author} {\bibfnamefont {K.-Y.}\ \bibnamefont
  {Yang}}, \bibinfo {author} {\bibfnamefont {Y.-M.}\ \bibnamefont {Lu}}, \ and\
  \bibinfo {author} {\bibfnamefont {Y.}~\bibnamefont {Ran}},\ }\href
  {https://journals.aps.org/prb/abstract/10.1103/PhysRevB.84.075129} {\bibfield
   {journal} {\bibinfo  {journal} {Physical Review B}\ }\textbf {\bibinfo
  {volume} {84}},\ \bibinfo {pages} {075129} (\bibinfo {year}
  {2011})}\BibitemShut {NoStop}%
\bibitem [{\citenamefont {Yang}\ \emph
  {et~al.}(2015{\natexlab{b}})\citenamefont {Yang}, \citenamefont {Pan},\ and\
  \citenamefont {Zhang}}]{yang2015chirality}%
  \BibitemOpen
  \bibfield  {author} {\bibinfo {author} {\bibfnamefont {S.~A.}\ \bibnamefont
  {Yang}}, \bibinfo {author} {\bibfnamefont {H.}~\bibnamefont {Pan}}, \ and\
  \bibinfo {author} {\bibfnamefont {F.}~\bibnamefont {Zhang}},\ }\href
  {https://journals.aps.org/prl/abstract/10.1103/PhysRevLett.115.156603}
  {\bibfield  {journal} {\bibinfo  {journal} {Physical review letters}\
  }\textbf {\bibinfo {volume} {115}},\ \bibinfo {pages} {156603} (\bibinfo
  {year} {2015}{\natexlab{b}})}\BibitemShut {NoStop}%
\bibitem [{\citenamefont {Cho}\ \emph {et~al.}(2012)\citenamefont {Cho},
  \citenamefont {Bardarson}, \citenamefont {Lu},\ and\ \citenamefont
  {Moore}}]{cho2012superconductivity}%
  \BibitemOpen
  \bibfield  {author} {\bibinfo {author} {\bibfnamefont {G.~Y.}\ \bibnamefont
  {Cho}}, \bibinfo {author} {\bibfnamefont {J.~H.}\ \bibnamefont {Bardarson}},
  \bibinfo {author} {\bibfnamefont {Y.-M.}\ \bibnamefont {Lu}}, \ and\ \bibinfo
  {author} {\bibfnamefont {J.~E.}\ \bibnamefont {Moore}},\ }\href
  {https://journals.aps.org/prb/abstract/10.1103/PhysRevB.86.214514} {\bibfield
   {journal} {\bibinfo  {journal} {Physical Review B}\ }\textbf {\bibinfo
  {volume} {86}},\ \bibinfo {pages} {214514} (\bibinfo {year}
  {2012})}\BibitemShut {NoStop}%
\bibitem [{\citenamefont {Wei}\ \emph {et~al.}(2014)\citenamefont {Wei},
  \citenamefont {Chao},\ and\ \citenamefont {Aji}}]{wei2014odd}%
  \BibitemOpen
  \bibfield  {author} {\bibinfo {author} {\bibfnamefont {H.}~\bibnamefont
  {Wei}}, \bibinfo {author} {\bibfnamefont {S.-P.}\ \bibnamefont {Chao}}, \
  and\ \bibinfo {author} {\bibfnamefont {V.}~\bibnamefont {Aji}},\ }\href
  {https://journals.aps.org/prb/abstract/10.1103/PhysRevB.89.014506} {\bibfield
   {journal} {\bibinfo  {journal} {Physical Review B}\ }\textbf {\bibinfo
  {volume} {89}},\ \bibinfo {pages} {014506} (\bibinfo {year}
  {2014})}\BibitemShut {NoStop}%
\bibitem [{\citenamefont {Kim}\ \emph {et~al.}(2016)\citenamefont {Kim},
  \citenamefont {Park},\ and\ \citenamefont {Gilbert}}]{kim2016probing}%
  \BibitemOpen
  \bibfield  {author} {\bibinfo {author} {\bibfnamefont {Y.}~\bibnamefont
  {Kim}}, \bibinfo {author} {\bibfnamefont {M.~J.}\ \bibnamefont {Park}}, \
  and\ \bibinfo {author} {\bibfnamefont {M.~J.}\ \bibnamefont {Gilbert}},\
  }\href {https://journals.aps.org/prb/abstract/10.1103/PhysRevB.93.214511}
  {\bibfield  {journal} {\bibinfo  {journal} {Physical Review B}\ }\textbf
  {\bibinfo {volume} {93}},\ \bibinfo {pages} {214511} (\bibinfo {year}
  {2016})}\BibitemShut {NoStop}%
\bibitem [{\citenamefont {Volovik}(2016)}]{Volovik2016}%
  \BibitemOpen
  \bibfield  {author} {\bibinfo {author} {\bibfnamefont {G.~E.}\ \bibnamefont
  {Volovik}},\ }\href {\doibase 10.1134/s0021364016210050} {\bibfield
  {journal} {\bibinfo  {journal} {{JETP} Letters}\ }\textbf {\bibinfo {volume}
  {104}},\ \bibinfo {pages} {645} (\bibinfo {year} {2016})}\BibitemShut
  {NoStop}%
\bibitem [{\citenamefont {Nissinen}\ and\ \citenamefont
  {Volovik}(2017)}]{Nissinen2017}%
  \BibitemOpen
  \bibfield  {author} {\bibinfo {author} {\bibfnamefont {J.}~\bibnamefont
  {Nissinen}}\ and\ \bibinfo {author} {\bibfnamefont {G.~E.}\ \bibnamefont
  {Volovik}},\ }\href {\doibase 10.1134/s0021364017070013} {\bibfield
  {journal} {\bibinfo  {journal} {{JETP} Letters}\ }\textbf {\bibinfo {volume}
  {105}},\ \bibinfo {pages} {447} (\bibinfo {year} {2017})}\BibitemShut
  {NoStop}%
\bibitem [{\citenamefont {Farajollahpour}\ \emph {et~al.}(2019)\citenamefont
  {Farajollahpour}, \citenamefont {Faraei},\ and\ \citenamefont
  {Jafari}}]{Tohid2019Spacetime}%
  \BibitemOpen
  \bibfield  {author} {\bibinfo {author} {\bibfnamefont {T.}~\bibnamefont
  {Farajollahpour}}, \bibinfo {author} {\bibfnamefont {Z.}~\bibnamefont
  {Faraei}}, \ and\ \bibinfo {author} {\bibfnamefont {S.~A.}\ \bibnamefont
  {Jafari}},\ }\href {\doibase 10.1103/PhysRevB.99.235150} {\bibfield
  {journal} {\bibinfo  {journal} {Phys. Rev. B}\ }\textbf {\bibinfo {volume}
  {99}},\ \bibinfo {pages} {235150} (\bibinfo {year} {2019})}\BibitemShut
  {NoStop}%
\bibitem [{\citenamefont {Weststr\"om}\ and\ \citenamefont
  {Ojanen}(2017)}]{OjanenPRX}%
  \BibitemOpen
  \bibfield  {author} {\bibinfo {author} {\bibfnamefont {A.}~\bibnamefont
  {Weststr\"om}}\ and\ \bibinfo {author} {\bibfnamefont {T.}~\bibnamefont
  {Ojanen}},\ }\href {\doibase 10.1103/PhysRevX.7.041026} {\bibfield  {journal}
  {\bibinfo  {journal} {Phys. Rev. X}\ }\textbf {\bibinfo {volume} {7}},\
  \bibinfo {pages} {041026} (\bibinfo {year} {2017})}\BibitemShut {NoStop}%
\bibitem [{\citenamefont {Liang}\ and\ \citenamefont
  {Ojanen}(2019)}]{Ojanen2019}%
  \BibitemOpen
  \bibfield  {author} {\bibinfo {author} {\bibfnamefont {L.}~\bibnamefont
  {Liang}}\ and\ \bibinfo {author} {\bibfnamefont {T.}~\bibnamefont {Ojanen}},\
  }\href {\doibase 10.1103/PhysRevResearch.1.032006} {\bibfield  {journal}
  {\bibinfo  {journal} {Phys. Rev. Res.}\ }\textbf {\bibinfo {volume} {1}},\
  \bibinfo {pages} {032006} (\bibinfo {year} {2019})}\BibitemShut {NoStop}%
\bibitem [{\citenamefont {Jalali-Mola}\ and\ \citenamefont
  {Jafari}(2019)}]{Sahar2019Polarization}%
  \BibitemOpen
  \bibfield  {author} {\bibinfo {author} {\bibfnamefont {Z.}~\bibnamefont
  {Jalali-Mola}}\ and\ \bibinfo {author} {\bibfnamefont {S.~A.}\ \bibnamefont
  {Jafari}},\ }\href {\doibase 10.1103/PhysRevB.100.075113} {\bibfield
  {journal} {\bibinfo  {journal} {Phys. Rev. B}\ }\textbf {\bibinfo {volume}
  {100}},\ \bibinfo {pages} {075113} (\bibinfo {year} {2019})}\BibitemShut
  {NoStop}%
\bibitem [{\citenamefont {Jafari}(2019)}]{Jafari2019}%
  \BibitemOpen
  \bibfield  {author} {\bibinfo {author} {\bibfnamefont {S.~A.}\ \bibnamefont
  {Jafari}},\ }\href {\doibase 10.1103/physrevb.100.045144} {\bibfield
  {journal} {\bibinfo  {journal} {Physical Review B}\ }\textbf {\bibinfo
  {volume} {100}} (\bibinfo {year} {2019}),\
  10.1103/physrevb.100.045144}\BibitemShut {NoStop}%
\bibitem [{\citenamefont {Kedem}\ \emph {et~al.}(2020)\citenamefont {Kedem},
  \citenamefont {Bergholtz},\ and\ \citenamefont {Wilczek}}]{Bergholtz2020}%
  \BibitemOpen
  \bibfield  {author} {\bibinfo {author} {\bibfnamefont {Y.}~\bibnamefont
  {Kedem}}, \bibinfo {author} {\bibfnamefont {E.~J.}\ \bibnamefont
  {Bergholtz}}, \ and\ \bibinfo {author} {\bibfnamefont {F.}~\bibnamefont
  {Wilczek}},\ }\href@noop {} {\bibfield  {journal} {\bibinfo  {journal} {arXiv
  preprint arXiv:2001.02625}\ } (\bibinfo {year} {2020})}\BibitemShut {NoStop}%
\bibitem [{\citenamefont {Soluyanov}\ \emph {et~al.}(2015)\citenamefont
  {Soluyanov}, \citenamefont {Gresch}, \citenamefont {Wang}, \citenamefont
  {Wu}, \citenamefont {Troyer}, \citenamefont {Dai},\ and\ \citenamefont
  {Bernevig}}]{Soluyanov2015}%
  \BibitemOpen
  \bibfield  {author} {\bibinfo {author} {\bibfnamefont {A.~A.}\ \bibnamefont
  {Soluyanov}}, \bibinfo {author} {\bibfnamefont {D.}~\bibnamefont {Gresch}},
  \bibinfo {author} {\bibfnamefont {Z.}~\bibnamefont {Wang}}, \bibinfo {author}
  {\bibfnamefont {Q.}~\bibnamefont {Wu}}, \bibinfo {author} {\bibfnamefont
  {M.}~\bibnamefont {Troyer}}, \bibinfo {author} {\bibfnamefont
  {X.}~\bibnamefont {Dai}}, \ and\ \bibinfo {author} {\bibfnamefont {B.~A.}\
  \bibnamefont {Bernevig}},\ }\href {\doibase 10.1038/nature15768} {\bibfield
  {journal} {\bibinfo  {journal} {Nature}\ }\textbf {\bibinfo {volume} {527}},\
  \bibinfo {pages} {495} (\bibinfo {year} {2015})}\BibitemShut {NoStop}%
\bibitem [{\citenamefont {Slichter}(2013)}]{slichter2013principles}%
  \BibitemOpen
  \bibfield  {author} {\bibinfo {author} {\bibfnamefont {C.~P.}\ \bibnamefont
  {Slichter}},\ }\href@noop {} {\emph {\bibinfo {title} {Principles of magnetic
  resonance}}},\ Vol.~\bibinfo {volume} {1}\ (\bibinfo  {publisher} {Springer
  Science \& Business Media},\ \bibinfo {year} {2013})\BibitemShut {NoStop}%
\bibitem [{\citenamefont {Alloul}(2015)}]{alloul2015nmr}%
  \BibitemOpen
  \bibfield  {author} {\bibinfo {author} {\bibfnamefont {H.}~\bibnamefont
  {Alloul}},\ }\href {https://arxiv.org/abs/1504.06992} {\bibfield  {journal}
  {\bibinfo  {journal} {arXiv preprint arXiv:1504.06992}\ } (\bibinfo {year}
  {2015})}\BibitemShut {NoStop}%
\bibitem [{\citenamefont {Okv{\'a}tovity}\ \emph {et~al.}(2016)\citenamefont
  {Okv{\'a}tovity}, \citenamefont {Simon},\ and\ \citenamefont
  {D{\'o}ra}}]{okvatovity2016anomalous}%
  \BibitemOpen
  \bibfield  {author} {\bibinfo {author} {\bibfnamefont {Z.}~\bibnamefont
  {Okv{\'a}tovity}}, \bibinfo {author} {\bibfnamefont {F.}~\bibnamefont
  {Simon}}, \ and\ \bibinfo {author} {\bibfnamefont {B.}~\bibnamefont
  {D{\'o}ra}},\ }\href
  {https://journals.aps.org/prb/abstract/10.1103/PhysRevB.94.245141} {\bibfield
   {journal} {\bibinfo  {journal} {Physical Review B}\ }\textbf {\bibinfo
  {volume} {94}},\ \bibinfo {pages} {245141} (\bibinfo {year}
  {2016})}\BibitemShut {NoStop}%
\bibitem [{\citenamefont {Yasuoka}\ \emph {et~al.}(2017)\citenamefont
  {Yasuoka}, \citenamefont {Kubo}, \citenamefont {Kishimoto}, \citenamefont
  {Kasinathan}, \citenamefont {Schmidt}, \citenamefont {Yan}, \citenamefont
  {Zhang}, \citenamefont {Tou}, \citenamefont {Felser}, \citenamefont
  {Mackenzie} \emph {et~al.}}]{yasuoka2017emergent}%
  \BibitemOpen
  \bibfield  {author} {\bibinfo {author} {\bibfnamefont {H.}~\bibnamefont
  {Yasuoka}}, \bibinfo {author} {\bibfnamefont {T.}~\bibnamefont {Kubo}},
  \bibinfo {author} {\bibfnamefont {Y.}~\bibnamefont {Kishimoto}}, \bibinfo
  {author} {\bibfnamefont {D.}~\bibnamefont {Kasinathan}}, \bibinfo {author}
  {\bibfnamefont {M.}~\bibnamefont {Schmidt}}, \bibinfo {author} {\bibfnamefont
  {B.}~\bibnamefont {Yan}}, \bibinfo {author} {\bibfnamefont {Y.}~\bibnamefont
  {Zhang}}, \bibinfo {author} {\bibfnamefont {H.}~\bibnamefont {Tou}}, \bibinfo
  {author} {\bibfnamefont {C.}~\bibnamefont {Felser}}, \bibinfo {author}
  {\bibfnamefont {A.}~\bibnamefont {Mackenzie}},  \emph {et~al.},\ }\href
  {https://journals.aps.org/prl/abstract/10.1103/PhysRevLett.118.236403}
  {\bibfield  {journal} {\bibinfo  {journal} {Physical review letters}\
  }\textbf {\bibinfo {volume} {118}},\ \bibinfo {pages} {236403} (\bibinfo
  {year} {2017})}\BibitemShut {NoStop}%
\bibitem [{\citenamefont {Okv{\'a}tovity}\ \emph {et~al.}(2019)\citenamefont
  {Okv{\'a}tovity}, \citenamefont {Yasuoka}, \citenamefont {Baenitz},
  \citenamefont {Simon},\ and\ \citenamefont
  {D{\'o}ra}}]{okvatovity2019nuclear}%
  \BibitemOpen
  \bibfield  {author} {\bibinfo {author} {\bibfnamefont {Z.}~\bibnamefont
  {Okv{\'a}tovity}}, \bibinfo {author} {\bibfnamefont {H.}~\bibnamefont
  {Yasuoka}}, \bibinfo {author} {\bibfnamefont {M.}~\bibnamefont {Baenitz}},
  \bibinfo {author} {\bibfnamefont {F.}~\bibnamefont {Simon}}, \ and\ \bibinfo
  {author} {\bibfnamefont {B.}~\bibnamefont {D{\'o}ra}},\ }\href
  {https://journals.aps.org/prb/abstract/10.1103/PhysRevB.99.115107} {\bibfield
   {journal} {\bibinfo  {journal} {Physical Review B}\ }\textbf {\bibinfo
  {volume} {99}},\ \bibinfo {pages} {115107} (\bibinfo {year}
  {2019})}\BibitemShut {NoStop}%
\bibitem [{\citenamefont {Maebashi}\ \emph {et~al.}(2019)\citenamefont
  {Maebashi}, \citenamefont {Hirosawa}, \citenamefont {Ogata},\ and\
  \citenamefont {Fukuyama}}]{maebashi2019nuclear}%
  \BibitemOpen
  \bibfield  {author} {\bibinfo {author} {\bibfnamefont {H.}~\bibnamefont
  {Maebashi}}, \bibinfo {author} {\bibfnamefont {T.}~\bibnamefont {Hirosawa}},
  \bibinfo {author} {\bibfnamefont {M.}~\bibnamefont {Ogata}}, \ and\ \bibinfo
  {author} {\bibfnamefont {H.}~\bibnamefont {Fukuyama}},\ }\href
  {https://www.sciencedirect.com/science/article/pii/S0022369717315755}
  {\bibfield  {journal} {\bibinfo  {journal} {Journal of Physics and Chemistry
  of Solids}\ }\textbf {\bibinfo {volume} {128}},\ \bibinfo {pages} {138}
  (\bibinfo {year} {2019})}\BibitemShut {NoStop}%
\bibitem [{\citenamefont {Hirata}\ \emph {et~al.}(2011)\citenamefont {Hirata},
  \citenamefont {Ishikawa}, \citenamefont {Miyagawa}, \citenamefont {Kanoda},\
  and\ \citenamefont {Tamura}}]{hirata201113}%
  \BibitemOpen
  \bibfield  {author} {\bibinfo {author} {\bibfnamefont {M.}~\bibnamefont
  {Hirata}}, \bibinfo {author} {\bibfnamefont {K.}~\bibnamefont {Ishikawa}},
  \bibinfo {author} {\bibfnamefont {K.}~\bibnamefont {Miyagawa}}, \bibinfo
  {author} {\bibfnamefont {K.}~\bibnamefont {Kanoda}}, \ and\ \bibinfo {author}
  {\bibfnamefont {M.}~\bibnamefont {Tamura}},\ }\href
  {https://journals.aps.org/prb/abstract/10.1103/PhysRevB.84.125133} {\bibfield
   {journal} {\bibinfo  {journal} {Physical Review B}\ }\textbf {\bibinfo
  {volume} {84}},\ \bibinfo {pages} {125133} (\bibinfo {year}
  {2011})}\BibitemShut {NoStop}%
\bibitem [{\citenamefont {Goerbig}\ \emph {et~al.}(2008)\citenamefont
  {Goerbig}, \citenamefont {Fuchs}, \citenamefont {Montambaux},\ and\
  \citenamefont {Pi{\'e}chon}}]{goerbig2008tilted}%
  \BibitemOpen
  \bibfield  {author} {\bibinfo {author} {\bibfnamefont {M.}~\bibnamefont
  {Goerbig}}, \bibinfo {author} {\bibfnamefont {J.-N.}\ \bibnamefont {Fuchs}},
  \bibinfo {author} {\bibfnamefont {G.}~\bibnamefont {Montambaux}}, \ and\
  \bibinfo {author} {\bibfnamefont {F.}~\bibnamefont {Pi{\'e}chon}},\
  }\href@noop {} {\bibfield  {journal} {\bibinfo  {journal} {Physical Review
  B}\ }\textbf {\bibinfo {volume} {78}},\ \bibinfo {pages} {045415} (\bibinfo
  {year} {2008})}\BibitemShut {NoStop}%
\bibitem [{\citenamefont {Trescher}\ \emph {et~al.}(2015)\citenamefont
  {Trescher}, \citenamefont {Sbierski}, \citenamefont {Brouwer},\ and\
  \citenamefont {Bergholtz}}]{trescher2015quantum}%
  \BibitemOpen
  \bibfield  {author} {\bibinfo {author} {\bibfnamefont {M.}~\bibnamefont
  {Trescher}}, \bibinfo {author} {\bibfnamefont {B.}~\bibnamefont {Sbierski}},
  \bibinfo {author} {\bibfnamefont {P.~W.}\ \bibnamefont {Brouwer}}, \ and\
  \bibinfo {author} {\bibfnamefont {E.~J.}\ \bibnamefont {Bergholtz}},\ }\href
  {https://journals.aps.org/prb/abstract/10.1103/PhysRevB.91.115135} {\bibfield
   {journal} {\bibinfo  {journal} {Physical Review B}\ }\textbf {\bibinfo
  {volume} {91}},\ \bibinfo {pages} {115135} (\bibinfo {year}
  {2015})}\BibitemShut {NoStop}%
\end{thebibliography}%


%






\end{document}